\documentclass[runningheads]{llncs}
\usepackage[T1]{fontenc}
\usepackage{graphicx}
\usepackage{hyperref}
\usepackage{xcolor}
\usepackage{todonotes}
\begin{document}
\title{AI Misuse in Education Is a Measurement Problem: Toward a Learning Visibility Framework}
\titlerunning{Ai Misuse in Education is a Measurement Problem}
%
\author{Eduardo Davalos\inst{1}\orcidID{0000-0001-7190-7273} \and
Yike Zhang\inst{2}\orcidID{0000-0003-3503-2996}}
%
\authorrunning{Davalos et al.}
%
\institute{
Trinity University, San Antonio TX, 78209, USA\\
\email{davalosedu@trinity.edu}\\
\and
St. Mary's University, San Antonio TX, 78228, USA\\
\email{yzhang5@stmarytx.edu}
}
\maketitle              
%


\begin{abstract}
The rapid integration of conversational AI systems into educational settings has intensified ethical concerns about academic integrity, fairness, and students' cognitive development. Institutional responses have largely centered on AI detection tools and restrictive policies, yet such approaches have proven unreliable and ethically contentious. This paper reframes AI misuse in education not primarily as a detection problem, but as a measurement problem rooted in the loss of visibility into the learning process. When AI enters the assessment loop, educators often retain access to final outputs but lose valuable insight into how those outputs were produced.
Drawing on research in cognitive offloading, learning analytics, and multimodal timeline reconstruction, we propose the \textbf{Learning Visibility Framework}, grounded in three principles: clear specification and modeling of acceptable AI use, recognition of learning processes as assessable evidence alongside outcomes, and the establishment of transparent timelines of student activity. Rather than promoting surveillance, the framework emphasizes transparency and shared evidence as foundations for ethical AI integration in classroom settings. By shifting focus from adversarial detection toward process visibility, this work offers a principled pathway for aligning AI use with educational values while preserving trust and transparency between students and educators.

 \\

\end{abstract}
\section{Introduction}

The release of ChatGPT in 2021 marked a turning point in education. Suddenly, the rapid adoption of the technology created widespread uncertainty among students and educators regarding the near-immediate transformation of teaching, learning, and assessment practices. Although early studies have begun to examine both short-term and long-term effects of conversational AI tools in educational contexts \cite{beckingham2024using,cukurova_artificial_2019,davalos_llms_2025,davar_ai_2025}, it remains premature to determine the full extent of their impact on student learning and academic development.

In the early stages of this technological shift, many educators turned to AI detection software, similar in spirit to plagiarism detection systems, in an effort to preserve academic integrity \cite{njee2025online}. However, these tools introduced new ethical and practical concerns. False positives led to disputes, reputation harm, and staining tensions between students and instructors \cite{Gorichanaz2023,King2025}. Beyond raising questionable accusations of academic misconduct, the limitations and inaccuracies of AI detection systems erode trust in institutional standing. It became increasingly clear that AI detection tools are neither sufficiently reliable nor ethically robust for high-stakes academic misconduct determinations. \cite{Deep2025}.

\begin{figure}
    \centering
    \includegraphics[width=\linewidth]{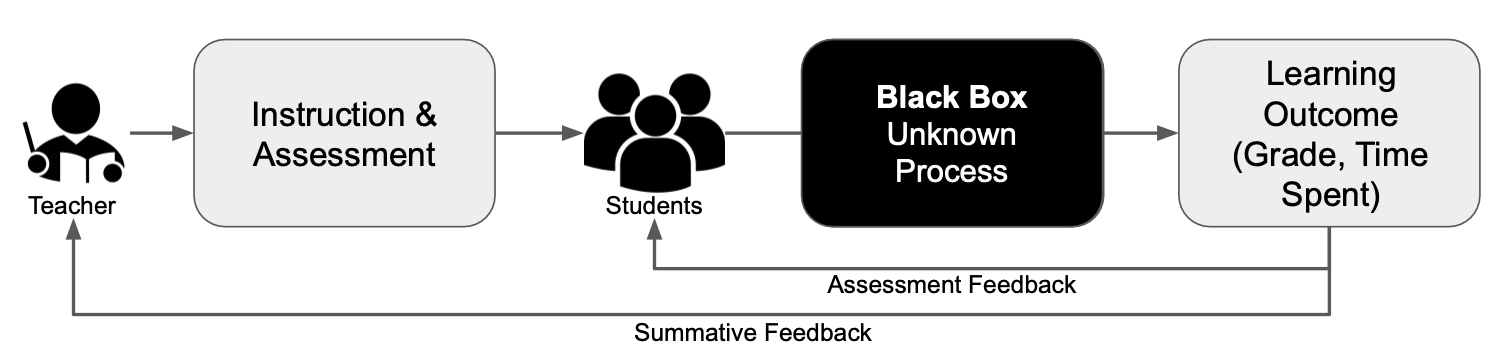}
    \caption{\textbf{Learning Visibility Problem}: When assessment relies primarily on observable outcomes such as grades or time spent, the underlying learning process remains opaque. The student’s cognitive and meta-cognitive activity becomes a ``black box,'' limiting educators’ ability to distinguish productive AI-supported learning from harmful cognitive offloading.}
    \label{fig:learning_visibility}
\end{figure}

Despite the shortcomings of many detection-based approaches, educators have limited alternatives for spotting and addressing AI misuse among students. Without effective mechanisms for observing and guiding students’ AI use, teachers face difficulty in helping students develop responsible, self-regulated learning practices alongside AI tools. As illustrated in Fig.~\ref{fig:learning_visibility}, the core issue is not merely detection, but visibility of the learning process. When AI enters the assessment process, learning becomes opaque. Educators lose valuable insight into how students develop their final answers, making it difficult to distinguish productive AI-supported learning from harmful cognitive offloading.
The absence of reliable evidence regarding students’ learning processes has broader consequences. Instructors may suspect AI misuse but lack sufficient documentation or evidence to uphold academic integrity policies. This situation enables unchecked and unregulated misuse in the classroom setting, where low-quality AI-generated outputs are submitted without meaningful engagement. Furthermore, when AI-generated work receives equal or higher grades than human-produced work, a feedback loop emerges that encourages greater reliance on AI tools. Over time, this dynamic risks reshaping academic norms to privilege output over understanding.

In this paper, we focus specifically on a form of AI misuse characterized by cognitive offloading through conversational agents. Students may delegate substantial portions of their coursework to AI systems and submit generated outputs with minimal oversight, reflection, or verification. Prior research on cognitive offloading suggests that excessive reliance on external systems can lead to shallow understanding and reduced opportunities to engage in higher-order cognitive and meta-cognitive processes \cite{Moqbel_2025,gerlich2025,fitriani2026academic}. Experimental studies have shown that when AI support is removed, students who previously relied heavily on such tools may struggle and perform worse than control groups without AI assistance \cite{Doleck2025}. These findings raise concerns about long-term skill development and conceptual mastery in course materials.

Yet, AI use is not inherently detrimental. AI tools can provide scaffolding, feedback, and support when used appropriately. The critical distinction lies in whether AI supplements cognition or replaces it. Ideally, AI should support meta-cognitive processes by encouraging reflection, planning, and evaluation, rather than substituting for core cognitive effort. However, ensuring such constructive use presents significant challenges in pedagogical design and development.
The proliferation of AI tools across digital learning environments further complicates the situation. Many platforms integrate AI assistance directly into workflows, and disabling these features is often impractical or not feasible. Additionally, the broader transition from paper-based to browser-based assessments has increased the accessibility and temptation of AI misuse. In response, some educators have reverted to in-person paper examinations using bluebooks to minimize AI interference. While this strategy reduces immediate risks, it reintroduces logistical burdens, environmental costs, and grading inefficiencies.

Rather than simply framing the problem solely as one of detection or prohibition, we argue that the \textit{AI misuse is fundamentally a measurement problem}. The central challenge is how to make learning processes visible in AI-assisted environments. To address this issue, we propose a visibility-focused framework grounded in three principles (\textbf{P1-3}) designed to support ethical, transparent, and pedagogically sound AI integration in classrooms.

\begin{itemize}
    \item \textbf{P1}: Clear specification, modeling, and disclosure of valid and invalid AI use in each assessment.
    \item \textbf{P2}: Recognition that both learning outcomes and learning processes, including student actions, revisions, and methods, are essential components of assessment.
    \item \textbf{P3}: Establishment of a transparent timeline of student actions and records to serve as shared evidence of learning.
\end{itemize}

In this paper, we elaborate on these three principles, relate them to existing pedagogical and technological practices, and discuss potential implementations to promote fairness, transparency, and responsible AI integration in education.

\section{Background}

This section reviews prior research on AI misuse in education, cognitive offloading, and the role of learning analytics in understanding the learning process. We draw from empirical studies across programming, writing, and neuroscience to situate the problem of AI-assisted learning within broader discussions of evidence, measurement, and pedagogical design.
Recent empirical work has examined how AI-assisted learning environments influence both students' performance and understanding. In a quasi-experimental study of 151 first-year computer science students, \cite{vivian2025coding} found substantial short-term gains in AI-assisted programming tasks, yet weak transfer to independent problem-solving contexts. These findings suggest that performance improvements do not necessarily translate into long-term learning.
Qualitative reflections indicated that students were often aware of gaps in their understanding but continued to rely heavily on AI systems. Drawing on Bloom’s taxonomy and cognitive offloading theory, the authors argue that sustained AI reliance may reduce higher-order cognitive engagement and algorithmic reasoning. Rather than portraying AI assistance as inherently harmful, the study calls for structured pedagogical integration that promotes meta-cognitive reflection and learner autonomy.

Similar findings have emerged in writing contexts. In studies of AI-assisted essay composition, \cite{ahmedtelba2025} observed that students who relied extensively on AI-generated text exhibited a superficial understanding of their own submissions. Participants often struggled to explain arguments, justify structural decisions, or recall specific claims within their essays. These outcomes suggest that excessive AI reliance may reduce opportunities for deep processing and overall knowledge construction.
Beyond behavioral and performance measures, emerging neuroscience-based research provides further insight into the cognitive effects of AI-assisted learning. In an electroencephalography study of essay writing, \cite{kosmyna2025brainchatgptaccumulationcognitive} compared participants using large language models (LLMs), search engines, and no external tools. Across sessions, LLM users exhibited weaker and less distributed brain connectivity than Brain-only participants, with cognitive activity scaling down in relation to external tool use. LLM users also reported lower ownership of their work and demonstrated difficulty recalling or accurately quoting their own essays.
Although such findings require careful interpretation, they suggest that sustained reliance on LLMs may be associated with reduced cognitive engagement and diminished internalization of knowledge. Together, these neural and behavioral patterns reinforce concerns that AI-mediated task completion can alter the intensity and distribution of cognitive effort during learning sessions.

\begin{figure}[h]
    \centering
    \includegraphics[width=\linewidth]{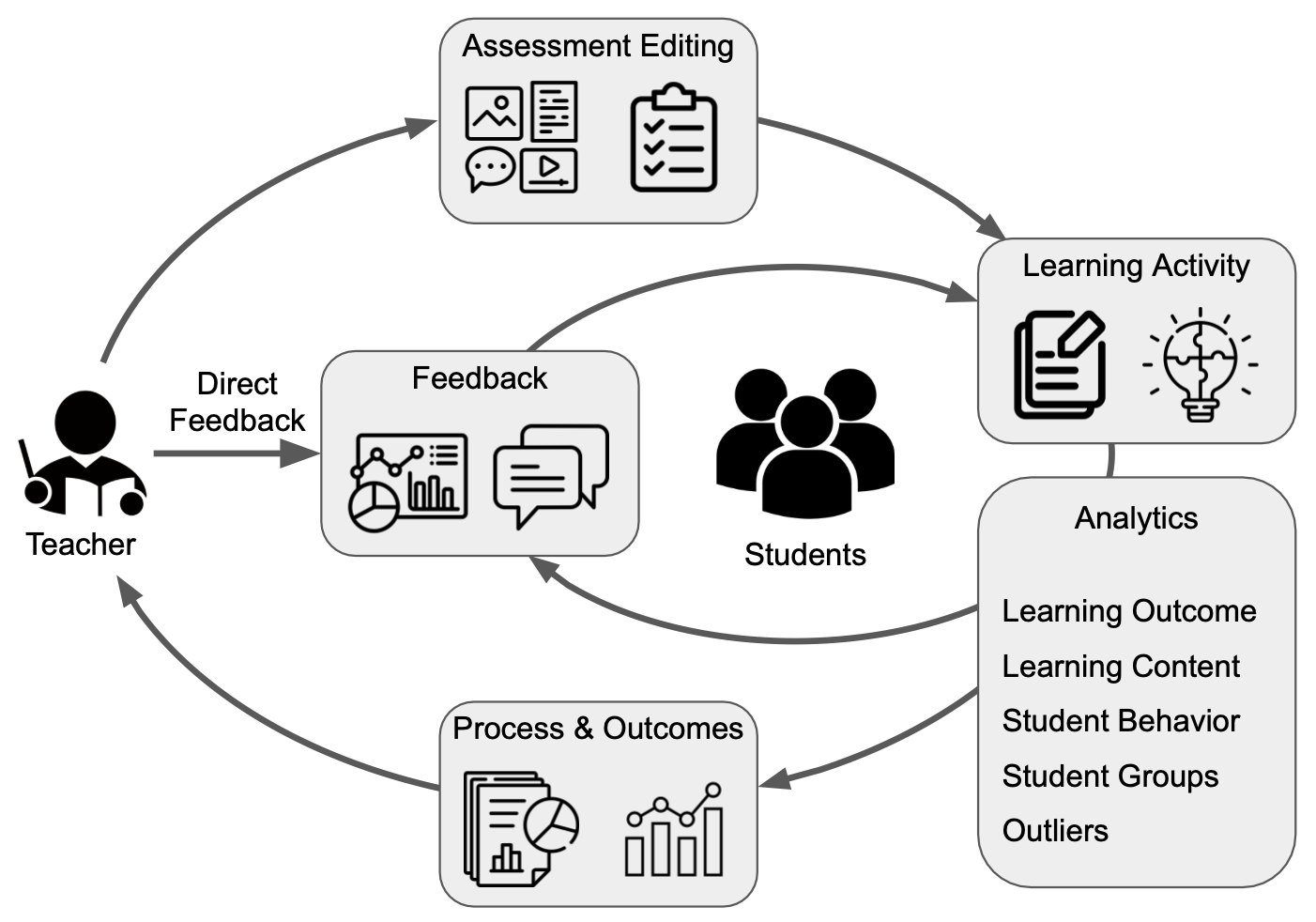}
    \caption{\textbf{Teacher and Student Feedback Cycles}: The relationship between these two interlinked cycles of intervention, analytics, and feedback. The inner cycle is composed of formative assessments while the outer cycle illustrates the summative cycle of learning analytics that aids teacher’s adaptive instruction and assessment.}
    \label{fig:research_framework}
\end{figure}

Together, these studies highlight the need to distinguish between ethical AI use and harmful cognitive substitution. They also underscore a central challenge for educators: determining how to observe and measure the quality of student engagement when AI tools are involved.
A critical component of monitoring AI use is the concept of showing evidence. In educational contexts, evidence consists of observable data points that substantiate claims about student understanding and effort. Such evidence can include data artifacts such as revision timestamps, intermediate problem-solving steps, code iteration histories, or interaction logs. When integrated with learning analytics techniques, these artifacts can provide insight into the processes that lead to a final submission.
The importance of integrating analytics into pedagogy has been emphasized in frameworks that position teachers as active interpreters of learning data. For example, \cite{wise_teaching_2019} describes a teaching with analytics model in which analytics are not merely descriptive dashboards but integral components of instructional decision-making. In this view, teachers gather, interpret, and act upon evidence generated through student interactions with learning systems. As illustrated in Fig.~\ref{fig:research_framework}, analytics-supported formative feedback cycles can inform summative assessment and adaptive instruction.
Despite this progress, many current learning analytics systems remain limited in scope. Commercial and institutional platforms often provide unimodal log-based indicators such as grades, time spent, and login frequency \cite{cohn2024multimodalmethodsanalyzinglearning}. While useful for identifying broad participation trends, these measures capture only a narrow portion of the learning process. They do not adequately reflect the complexity of cognitive, meta-cognitive, and social dynamics involved in authentic learning tasks.

To address these limitations, the field of multimodal learning analytics has emerged. Multimodal approaches integrate diverse data sources, including interaction logs \cite{cohn2024chain,munshi_adaptive_2022}, gaze tracking \cite{davalos_gazeviz_2024,davalos2023identifyinggraphs}, affect \cite{nkambou_predicting_2018}, posture \cite{mitri_detecting_nodate}, speech \cite{munoz_development_nodate}, physiological signals \cite{rau_integrating_2019}, and collaborative behaviors \cite{nasir_many_2021}, to build richer models of learner engagement and self-regulation \cite{ochoa2022multimodal,worsley_multimodal_2018,blikstein_multimodal_2016}. Compared to simple log data, multimodal systems have demonstrated improved sensitivity in identifying patterns of cognition, collaboration, and effort.
Importantly, multimodal learning analytics has also advanced the reconstruction and visualization of learning timelines \cite{echeverria_driving_2018}. Timeline-based representations organize heterogeneous data streams into coherent temporal narratives of student activity. For example, \cite{FONTELES2026} developed a timeline visualization tool for collaborative embodied STEM learning that combined system logs with social signal data such as gaze direction, posture, spatial position, and speech. Similarly, \cite{martinez-maldonado_data_2020} implemented timeline analytics in nursing simulation training, enabling instructors to review student nurses’ actions throughout complex scenarios and provide holistic, evidence-based feedback.

Across these contexts, timeline representations serve a common purpose: they transform dispersed interaction traces into interpretable evidence of human action. The resulting data artifacts are generated through students’ observable behaviors and therefore provide indirect insight into underlying cognitive and meta-cognitive processes. In AI-assisted learning environments, timeline-based evidence can help restore visibility into otherwise opaque learning processes.

\section{Learning Visibility Framework}

This section introduces the three core principles of the \textbf{Learning Visibility Framework} and explains why they are essential for communicating expectations, monitoring engagement, and regulating AI use in educational settings. Rather than focusing on enforcement or prohibition, the framework emphasizes transparency, shared understanding, and measurable evidence of learning processes.

\subsection*{P1: Clear Specification and Modeling of Valid and Invalid AI Use}

The first principle focuses on communication and transparency of expectations between students and educators. In the wake of widespread concerns about AI misuse and false accusations, maintaining bidirectional trust has become increasingly important. Ambiguity regarding acceptable AI use erodes confidence on both sides. When policies are unclear or inconsistently enforced, misunderstandings multiply and the teacher-student partnership deteriorates.
\begin{figure}[t]
    \centering
    \includegraphics[width=\linewidth]{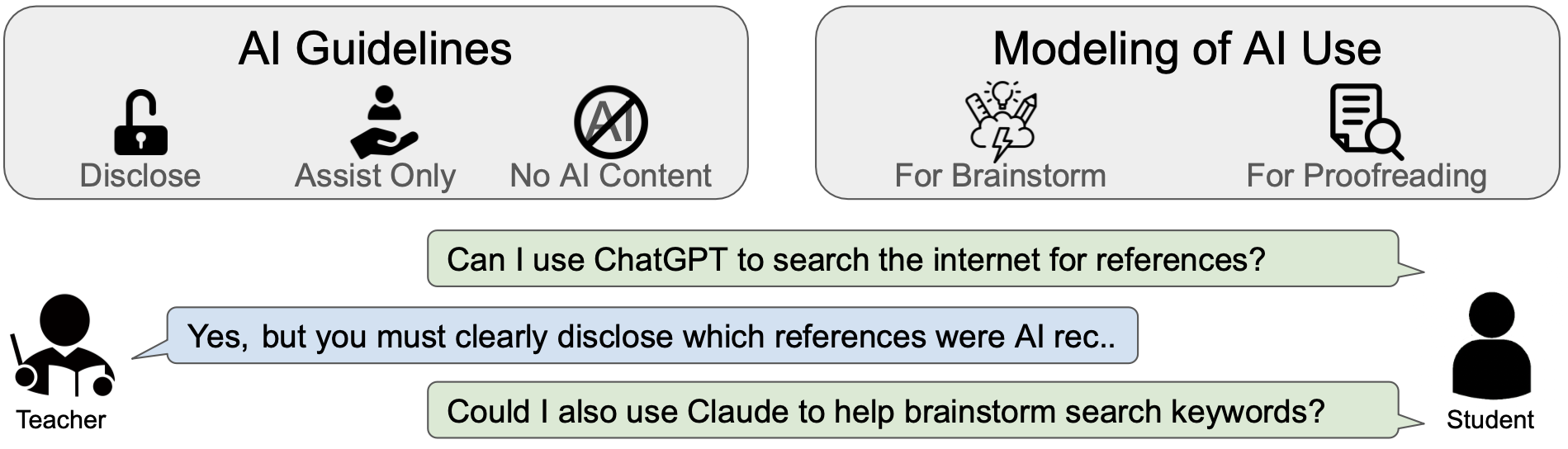}
    \caption{\textbf{P1: Clear Specification and Modeling of AI Use}. Explicit guidelines and example use cases, reinforced through open teacher–student dialogue, establish shared expectations and reduce ambiguity around acceptable AI use.}
    \label{fig:p1_illustration}
\end{figure}
From the educator’s perspective, it is essential to clearly define and model acceptable AI use within each assessment. Modeling extends beyond listing permitted or prohibited tools, as it includes demonstrating appropriate prompts, responsible workflows, and reflective practices that align with learning objectives. By explicitly stating valid and invalid uses of AI, instructors can reduce unintentional misuse and establish shared norms. As illustrated in Fig.~\ref{fig:p1_illustration}, clear guidelines combined with example use cases and open dialogue transform policy statements into shared expectations. The teacher–student exchange described in the figure emphasizes that transparency is not merely declarative, but conversational and iterative.

Effective modeling should center on cognitive and meta-cognitive development. AI systems should not remove human autonomy or replace essential cognitive effort. Instead, they should support planning, reflection, and self-evaluation. The appropriate boundary depends on the instructional context. In introductory programming assignments, where the primary objective is to develop foundational syntax understanding and problem-solving skills, instructors may prohibit AI-generated code entirely. In contrast, in writing-intensive courses, AI tools may be allowed for brainstorming or high-level feedback, as long as students produce and revise their own substantive work.

A notable challenge is that widely used conversational agents are frequently treated as tutors, yet they are not designed with pedagogical guardrails that reliably prevent over-scaffolding or solution disclosure. Without structured guidance, these systems may provide answers rather than promote content understanding. Human instructors can make similar mistakes when they offer complete solutions instead of scaffolding reasoning. The framework therefore emphasizes that modeling responsible AI use requires intentional instructional design, not merely access to technology.

\subsection*{P2: Valuing Both Learning Outcomes and Learning Processes}

The second principle shows that both outcomes and processes are essential components of assessment. Historically, formative and summative evaluations have relied primarily on final products such as completed exams, essays, projects, or problem sets. The finished artifact served as the primary measurement instrument for judging student proficiency and course materials mastery. However, AI systems can generate polished outputs that resemble genuine work, making it increasingly difficult to distinguish between authentic learning and delegated production based solely on final submissions.

\begin{figure}[t]
    \centering
    \includegraphics[width=\linewidth]{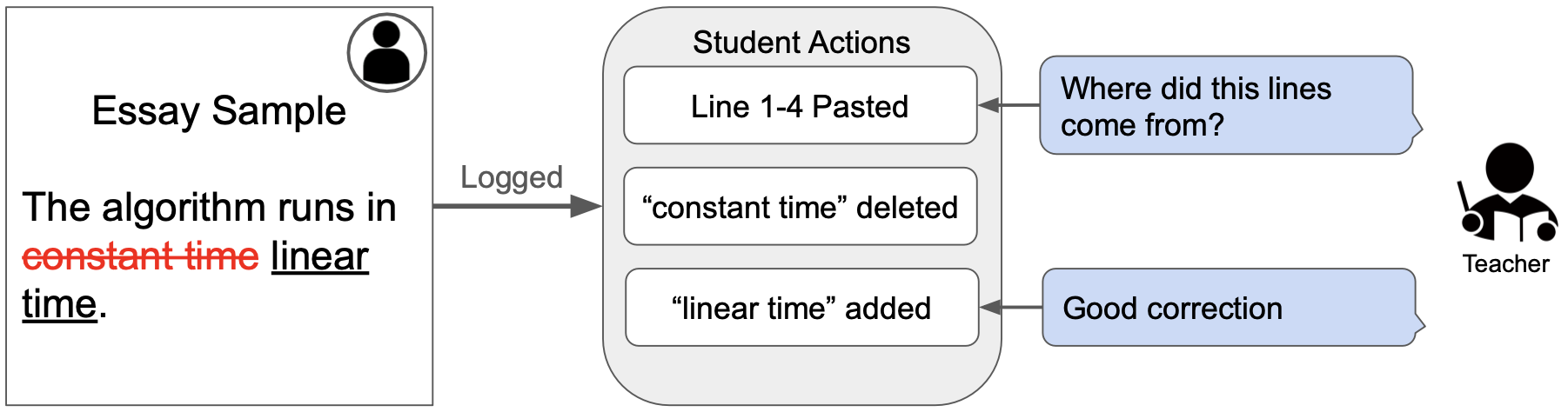}
    \caption{\textbf{P2: Learning Process as Evidence}. Revision traces, insertions, and deletions provide visible records of student activity that allow educators to interpret effort, reasoning, and potential AI involvement beyond the final submitted product.}
    \label{fig:p2_illustration}
\end{figure}

To address this limitation, assessment practices must shift from outcome-only evaluation toward incorporating process-based evidence. The learning process itself should be considered as a measurable and assessable dimension of student performance. Observable behaviors such as revision patterns, iteration histories, intermediate drafts, and problem-solving steps provide insight into engagement and effort. As illustrated in Fig.~\ref{fig:p2_illustration}, visible traces of edits, insertions, and deletions transform the essay from a static product into a record of evolving reasoning. Process data, such as essay revisions, enable educators to ask informed questions and distinguish constructive effort from AI delegation.
Some tools have begun to support this transition. For example, revision history tracking in collaborative writing platforms enables instructors to examine how students develop and refine their ideas over time \cite{Turkay2018}. These tools provide temporal records that contextualize final submissions. However, equivalent visibility mechanisms are less common in domains such as mathematics, computer science, physics, and other problem-solving disciplines. The absence of comprehensive process-tracking infrastructure limits the scalability of visibility-based assessment across curricula.

Process visibility serves not only as a deterrent to misuse but also as a pedagogical resource. Access to the learning process data can address misconceptions, inefficient strategies, and gaps in understanding. Importantly, the interpretation of such evidence should remain the responsibility of human educators. While analytics systems can summarize or visualize behavioral data, the ethical and contextual judgment required to identify misuse or academic dishonesty must not be delegated entirely to automated systems. Retaining educator oversight preserves fairness and reduces the risk of algorithmic mis-classification.

\subsection*{P3: Establishing a Transparent Timeline of Learning Activity}

\begin{figure}
    \centering
    \includegraphics[width=\linewidth]{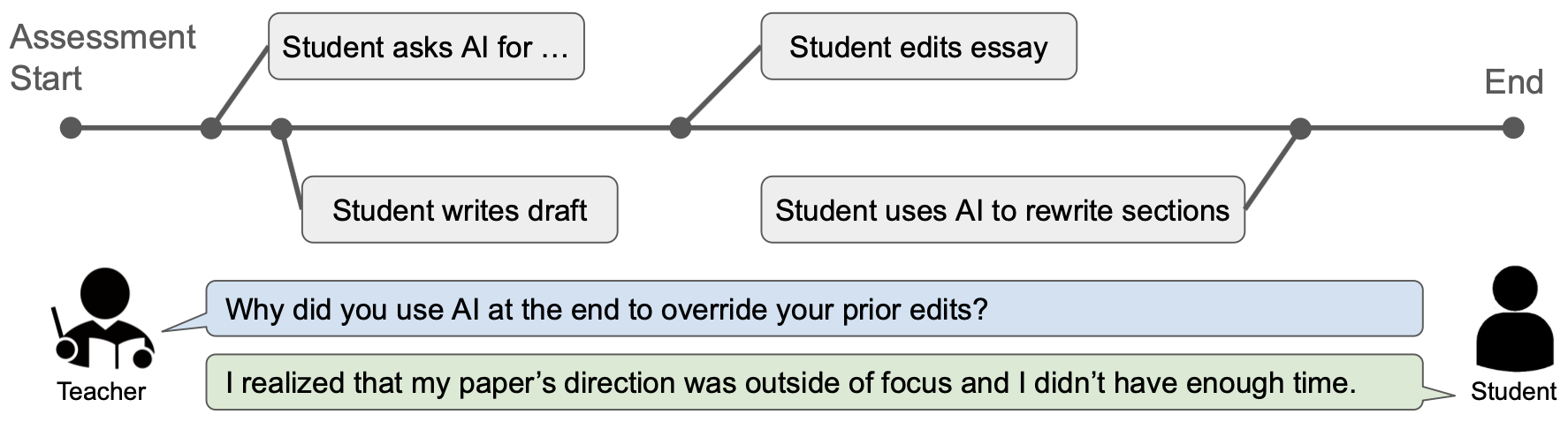}
    \caption{\textbf{P3: Transparent Timeline of Learning Activity}. A temporal record of student actions reveals sequences of drafting, AI interaction, and revision, enabling educators to contextualize AI use and engage students in reflective dialogue about their decision-making process.}
    \label{fig:p3_illustration}
\end{figure}

The third principle emphasizes the importance of constructing a transparent timeline of student activity as a shared artifact within the teacher-student partnership. A timeline organizes discrete interaction traces into a coherent temporal narrative of learning. This representation functions as both a record of engagement and a tool for dialogue. As shown in Fig.~\ref{fig:p3_illustration}, mapping student actions across time reveals how drafting, AI interaction, and revision unfold within a single assessment. This temporal structure enables educators to contextualize AI use rather than judge isolated edits or final submissions.

Timeline-based evidence aligns with developments in multimodal learning analytics, where diverse data sources are integrated to reconstruct sequences of cognitive and behavioral events \cite{worsley_situating_2016,fonteles2024}. Such approaches have demonstrated applicability across varied domains and learning environments. A structured timeline enables educators to observe patterns such as abrupt content insertion, prolonged inactivity followed by rapid completion, or iterative refinement consistent with sustained effort.

Beyond identifying misuse, the timeline supports formative feedback. By reviewing sequences of actions, instructors can pinpoint moments where misconceptions arise or strategies falter. Students, in turn, gain opportunities to reflect on their workflows and decision-making processes. The timeline thus becomes not merely a surveillance mechanism but a collaborative reference point for discussion and improvement.

\subsection*{Integrating the Three Principles}

Taken together, the three principles establish a coherent framework for addressing AI misuse without defaulting to adversarial detection systems. Clear communication and modeling of acceptable AI use provide normative guidance. Valuing learning processes alongside outcomes expands the evidentiary basis of assessment. Transparent timelines transform interaction traces into interpretable artifacts that support dialogue and overall accountability.
When implemented collectively, these principles aim to reduce distrust and promote fairness. Rather than framing AI use as inherently suspect, the framework positions visibility and shared evidence as the foundation for ethical integration. By making goals, processes, and expectations transparent, educators and students can move toward consensus-driven practices that assist learning while preserving academic integrity.

\section{Challenges, Limitations, and Future Work}

While the \textbf{Learning Visibility Framework} offers a structured approach to addressing AI misuse, several important challenges and limitations must be acknowledged.

A primary concern involves student privacy. Systems designed to capture and analyze process data necessarily collect detailed records of student actions during assessments. Without careful design, such data collection risks infringing upon student autonomy and confidentiality. Educational technology platforms and instructors must therefore transparently disclose what data are collected, how they are used, and who has access to them. Informed consent and clear disclosure mechanisms are essential. Moreover, stakeholders must adopt principles of data minimization, collecting only what is necessary to support pedagogical goals. Ethical implementation requires that process visibility not become disproportionate surveillance in practice. Protecting student safety and dignity must remain a foundational design principle and constraint.

A second challenge concerns the volume and complexity of process data. Detailed interaction traces can quickly accumulate into large and cognitively demanding datasets. Many existing learning analytics dashboards already present instructors with dense visualizations that are difficult to interpret holistically. If poorly designed, visibility systems may increase instructor workload rather than reduce uncertainty. Future work therefore should focus on human-centered interface design that supports efficient interpretation without cognitive overload. This includes prioritizing meaningful patterns over raw logs, providing layered summaries with drill-down capabilities, and aligning analytics displays with instructional decision-making processes. Effective visibility should enhance pedagogical insight rather than overwhelm educators with excessive detail.

A further limitation involves the inevitability of circumvention. Students who are highly motivated to evade detection may adapt their behavior to bypass visibility mechanisms. For instance, instead of copy-pasting AI-generated text, a student might manually transcribe it to create the appearance of authentic revision history. However, such adversarial strategies are not unique to AI-enabled environments. Academic dishonesty has long included external assistance such as paid writing services or unauthorized collaboration \cite{AMIGUD2019}. The proposed framework does not aim to eliminate all forms of deliberate cheating.

Overall, the framework primarily targets widespread, low-reflection misuse that arises from ambiguity, convenience, or lack of guidance. By emphasizing transparency, shared evidence, and pedagogical alignment, the goal is to reduce unintentional or habitual cognitive offloading. Future research should empirically evaluate whether visibility-based approaches can meaningfully guide student behavior, improve meta-cognitive engagement, and maintain trust while respecting ethical boundaries.
Taken together, these challenges highlight that visibility is not a purely technical solution. It is a socio-technical intervention that must balance privacy, usability, fairness, and instructional intent. Continued interdisciplinary collaboration among educators, researchers, and technology designers is necessary to refine and responsibly implement this framework.

\section{Conclusion}

The swift integration of conversational AI systems into educational settings has intensified concerns about academic integrity, cognitive development, and fairness. Initial responses have focused on detection and prohibition, yet these approaches have proven unreliable and ethically problematic. This paper reframes AI misuse not as a detection problem, but as a measurement problem. When AI enters the assessment loop, the central loss is visibility into the learning process.

Drawing on research in cognitive offloading and learning analytics, we argue that responsible AI integration requires advocating transparency to student activity. We introduce the \textbf{Learning Visibility Framework}, grounded in three principles: clear specification and modeling of acceptable AI use, recognition of learning processes as assessable evidence alongside outcomes, and the establishment of transparent timelines of student actions. Together, these principles shift the focus from adversarial enforcement toward shared evidence and pedagogical clarity.

Rather than attempting to eliminate all types of academic misconduct, the framework seeks to reduce ambiguity, support meta-cognitive engagement, and preserve trust between students and educators. Visibility is not a surveillance mechanism but a socio-technical strategy for aligning ethical AI use with long-standing educational values and well-grounded learning theory. As AI systems continue to evolve, institutions must move beyond reactive enforcement toward principled integration. Ultimately, the ethical use of AI in education depends not only on what students produce, but on how they actively learn.

\begin{credits}

\subsubsection{AI Use} ChatGPT was used for text editing and formatting of the manuscript. The authors have manually inspected, corrected, and verified the generated text and take full responsibility of the manuscript's claims and content.

\subsubsection{\discintname}
The authors have no competing interests to declare that are relevant to the content of this article. 
\end{credits}
%
%
%
\bibliographystyle{splncs04}
\bibliography{sorted_main,new_entries}

\end{document}